\documentstyle[12pt,epsfig]{article}
\begin{document}
\title{On a Quantum Equivalence Principle.}
\author{ A. Camacho
\thanks{email: acamacho@aip.de} \\
Astrophysikalisches Institut Potsdam. \\
An der Sternwarte 16, D--14482 Potsdam, Germany.}

\date{}
\maketitle

\begin{abstract}

The logical consistency of a description of Quantum Theory in the context of General Relativity, which includes Minimal 
Coupling Principle, is analyzed from the point of view of Feynman's formulation in terms of path integrals. 
We will argue from this standpoint and using an argument that claims the incompleteness 
of the general--relativistic description of gravitation, which emerges as a consequence of the gravitationally induced phases of the so called 
flavor--oscillation clocks, that the postulates of Quantum Theory are logically incompatible with the usual Minimal Coupling Principle. 
It will be shown that this inconsistency could emerge from the fact that the required geometrical information to calculate the probability 
of finding a particle in any point of the respective manifold does not lie in a region with finite volume. 
Afterwards, we put forth a new Quantum Minimal Coupling Principle in terms of a 
restricted path integral, 
and along the ideas of this model not only the propagator of a free particle is 
calculated but we also deduce the conditions under which we recover Feynman's case for a free particle. 
The effect on diatomic interstellar molecules is also calculated.
The already existing relation between Restricted Path Integral For\-ma\-lism\- and 
Decoherence Model will enable us to connect the issue of a Quantum Minimal Coupling 
Principle with the collapse of the wave function. From this last remark we will claim that the geometrical structure 
of the involved manifold acts as, an always present, measuring device on a quantum particle. 
In other words, in this proposal we connect the issue of a Quantum Minimal Coupling Principle with a claim which states that gravity could be one of the physical entities driving the collapse of the wave function.

\end{abstract}

\newpage
\section{Introduction.}
\bigskip

One of the most ambitious programs in Modern Physics comprises the quantization of General Relativity [1]. 
Nevertheless, the proposed solutions to this old conundrum have up to now, in one way or in another, failed [2].

The solutions that in this context already exist assume always the logical (at least in some limit case) 
consistency of a theory involving simultaneously the postulates of Quantum Theory (QT) 
and of General Relativity (GR). In connection with this issue, one of the most thorny points 
is related to the validity in QT of the Minimal Coupling Principle (MCP). 
It is true that there are claims which state that the usual MCP 
is valid even in QT [3, 4]. Neverwithstanding, recently 
some works have appeared which contemplate the possible inconsistency of MCP on the quantum level [5].

In this work we will analyze more carefully this last claim and try to give an answer 
to the following questions:

1) Which postulate(s) of GR can be held responsible for the aforementioned incompleteness?

2) Could we define a logically consistent Quantum Minimal Coupling Principle (QMCP)?

Concerning the first question we will give an argument which claims that this inconsistency stems 
from the non--local character of the information that is required in order to determine in every point 
of spacetime the corresponding wave function. This fact will be explained constructing 
two {\it Gedankenexperimente}. One of them will be the usual two--slit experiment with the whole measuring 
device situated in a region containing a nonvanishing gravitational field in such a way that the whole experimental 
apparatus lies completely within the validity region of a locally flat coor\-di\-nate system (LFCS). The second one 
will be based upon Ahluwalia's flavor--oscillation clock [5]. These two {\it Gedankenexperimente} will be analyzed 
using Feynman's formulation in terms of path integrals. As a consequence of MCP the experimental outputs of the first case imply that curvature 
has no effect on the contribution to the propagator of those trajectories which do not lie 
completely within the validity region of LFCS, whereas 
the second one asserts that curvature does affect this propagator. It will be shown that this 
inconsistency could emerge from the fact that the required geo\-me\-tri\-cal information to calculate the probability of finding a particle 
in any point of the respective manifold does not lie in a region with finite volume. 
This does not happen in classical me\-cha\-nics, where the required geometrical information to describe the 
movement between two points can always be enclosed in a region with finite volume.

Regarding the second question we will try to obtain a logically consistent des\-cription of QT 
in the context of GR by means of a new QMCP. 
The original idea here consists in the restriction of the integration domain 
of the involved path integral. In other words, this new QMCP is expressed in terms of the so called Restricted Path Integral Formalism (RPIF). 
Afterwards, we will calculate using this idea the case of a free particle and recover, as a limit situation, Feynman's propagator for a free particle. 
The physical conditions under which this happens are also obtained.

Finally, recalling the relation between RPIF and Decoherence Model (DM) we will be 
able to connect the issue of a logically consistent QMCP with the problem of the collapse of the wave function. 
In other words, with this proposal we could relate the topic of a QMCP with a claim 
stating that the gravitational field could be one of the physical entities driving the collapse of the wave function [6, 7].

At this point it is also important to add that there are already attempts to 
formulate a quantum equivalence principle in the context of Feynman's idea [8]. This has been 
done starting from a path integral in a flat space parametrized with euclidean 
coordinates. Afterwards, a non--holonomic coordinate transformation is carried out and in this way a path integral is obtained, and the involved coordinate transformation is claimed to be 
a quantum equivalence principle which allows us to generalize the Feynman path integral formula 
in cartesian coordinates to non--euclidean spaces [9]. 
Neverwithstanding, it is very important to distinguish between geometrical effects and accelerative effects [10]. 
If we had a flat space and transform in a non--holonomic ma\-nner\- its euclidean coordinates, then we go to an a\-cce\-le\-ra\-ted reference frame, 
but this transformation does not endow our initial flat manifold with nonvanishing curvature, 
i.e., there is still no gravitational field. 
What has then been obtained is the description in an accelerated reference frame of the corresponding path integral. 
But if we insist in interpreting the resulting path integral as the propagator in a curved manifold 
(as a consequence of the equivalence between gravity and accelerated frames), then we may convince ourselves very easily that 
this proposal can not render the correct path integral in an arbitrary curved manifold. We may understand this point better 
noting that this non--holonomic coordinate transformation ($dx^i \rightarrow c^i_{\mu}(q)dq^{\mu}$, where $c^i_{\mu}(q) = {\partial x^i\over \partial q^{\mu}}$ are the so called basis triads [9]) 
can be contemplated from a different point of view, namely we begin with a curved manifold in which it is possible 
to define a globally flat coordinate system. This last condition imposes a 
very stringent geometrical res\-tric\-tion on our curved manifold. Indeed, we already know that 
in the most general case this condition is not fulfilled, the presence of tidal forces allows the definition of locally flat coordinate systems but will not  in general allow the definition of a globally flat coordinate system. 
In other words, this quantum equivalence principle does not render the path integral of a particle in an arbitrary 
curved ma\-ni\-fold because from the very begining it assumes the absence of tidal forces, i.e., the absence of nonuniformities in the gravitational field.
\bigskip

\section{Path Integrals and Minimal Coupling Principle.}
\bigskip

To understand a little bit better this incompleteness argument [5], let us at this point 
consider an arbitrary nonvanishing gravitational field, and pick out a certain point $P$ in this manifold. 
The geometrical properties of GR allow us to define a LFCS, 
whose origin coincides with point $P$ and which is valid only for points ``sufficiently close'' 
to $P$. Take now a second point $A$, the only condition that we impose on this point 
is that it has to be located in the validity region of LFCS. 

At this point let us be a little more explicit about the meaning of the phrase ``validity region of the locally flat coordinate system''. 
A correct geometrical definition of LFCS is given by the so called Fermi Normal Coordinates [11], accurate to second order. 
The deviation from the flat case of the $g_{\mu\nu}$ term is proportional to $R_{\mu l\nu m}x^{l}x^{m}$, i.e., 
the local metric takes the form $g_{\mu\nu} = \eta_{\mu\nu} + \alpha R_{\mu l\nu m}x^{l}x^{m} + O(|x^j|^3)dx^{\alpha}dx^{\beta}$, 
here $| \alpha | \in [0, 4/3]$, $x^{l}$ are the local space--like Lorentz coordinates, and 
$R_{\mu l\nu m}$ are the components of the Riemann tensor along the world line $x^j = 0$. 
With this metric we may estimate the size of the validity region of LFCS, i.e., 
it comprises those points which satisfy the condition $|R_{\mu l\nu m}x^{l}x^{m}| << 1$. 

Assume now that a freely falling quantum mechanical particle was 
in point $P$, and let us ask for the probability of finding this particle in $A$. 
From MCP we have that in LFCS the particle is described 
by the free particle Schr\"odinger equation (here we will restrict ourselves to the analysis of the limit of low velocities of Dirac's equation). Up to this point everything seems to be logically 
consistent. But we know that Schr\"odinger formulation of QT is completely equivalent 
to Feynman's one [12]. Therefore we must be able to find this probability using Feynman's idea, 
otherwise we would have the breakdown of MCP. Thus, in this formulation, the probability of 
finding our particle in $A$ is constructed with the contribution of all 
the trajectories that join $P$ and $A$. At this very same point we face already a conceptual problem, 
namely in order to perform this sum (integration) we must consider not only trajectories 
which can be described by LFCS, but also trajectories beyond the validity region of this coordinate system. 
Therefore it seems that the perspective that we obtain from this issue employing Feynman's formulation 
could mean that LFCS and the laws of Physics in Special Relativity 
could not suffice to obtain a complete description of the aforementioned probability. 
It seems that this argumentation already supports 
the incompleteness of GR in the context of QT that has already been pointed out [5].

Of course, that if we do not have to sum (integrate) over all the possible trajectories and 
had to sum only over those trajectories which can be described by LFCS, then this inconsistency would disappear. Clearly, that would also require 
the introduction of a weight functional in Feynman's path integral formulation, 
otherwise we could not obtain, at least as a limit case, Feynman's result. We may understand better this 
point noting that if the components of the Riemann tensor increase then the validity 
region of LFCS becomes smaller, and therefore less trajectories will appear in the path integral. But we 
will expect to obtain under some conditions (for example, when the length of the classical trajectory is much smaller than the validity region of LFCS) Feynman's case, which considers all possible trajectories. This condition could be fulfilled with the 
introduction a weight functional.

At this point let us underline by means of two {\it Gedankenexperimente} the logical inconsistency of 
a theory containing the postulates of QT and the usual MCP.

Firstly, consider the usual two--slit experiment [13], but this time, let us also 
a\-ssu\-me that the whole experimental device is immersed in a region that contains a nonvanishing 
gravitational field such that this experimental apparatus lies completely inside 
the validity region of a locally flat coordinate system. Therefore the interference 
pattern that would appear in the corresponding detecting screen is, as a consequence of MCP, the same interference pattern 
that emerges in the case without gravitational field. This result 
implies that the contribution to the interference pattern, coming from those trajectories 
not lying completely in the validity region of the locally flat 
coordinate system, is the same as in the case in which we had no gravitational field.

At this point it is noteworthy to mention that this last {\it Gedankenexperiment} 
is not the experimental construction of Colella's {\it et al} [14]. Indeed, this experimental apparatus is not located within 
the validity region of a LFCS because in it the effect of gravity on the interference pattern of 
two neutron beams is analyzed.

Secondly, let us take up Ahluwalia's most important result [5]. He asserts that 
the frequency of a ``flavor--oscillation clock'' $\Omega^F$ [15] in a freely falling frame in Earth's gravity 
and the same frequency $\Omega^{\infty}$ in a gravity--free region satisfy 
the condition 
$\Omega^F < \Omega^{\infty}\Rightarrow 1/\Omega^{\infty} < 1/\Omega^F$. 
Such clocks are constructed as a quantum mechanical superposition of different mass eigenstates, for 
instance two neutrinos from different lepton generation, 
$|F_a > = cos(\theta)|m_1> + sin(\theta)|m_2>$ and 
$|F_b > = -sin(\theta)|m_1> + cos(\theta)|m_2>$.

In this last argument the gravitational system is composed by the Earth and the local cluster of 
galaxies, the so called Great Attractor. We must also comment that the aforementioned effect 
emerges because the gravitational potential of this system $\phi_{effe.}$ is 
for points near the Earth's surface given by two contributions 
$\phi_{effe.} = \phi_{E} + \phi_{GA}$. The first one $\phi_{E}$ stems from Earth's mass while the second one 
$\phi_{GA.}$ comes from the Great Attractor. This second term is constant up to one part in about $10^{11}$. 
Therefore if we go now to a freely falling reference frame near Earth's surface we may get rid 
(as a consequence of MCP) of all gradients of the gravitational potential, 
nevertheless its constant parts will survive, i.e., $\phi_{E}$ disappears but $\phi_{GA}$ 
is preserved. In other words, gravity--induced accelerations vanish but the 
constant parts of it have a physical effect (via $\phi_{GA}$--dependent gravitationally induced phases), something similar to the Aharanov--Bohm effect [16].

We may now measure time with these clocks, and thus if we consider the clock 
situated in the freely falling frame and assume that we started with flavor state $|F_a>$ 
and ask now for the probability of having flavor state $|F_b >$ at a proper time 
$\tau = 1/\Omega^{\infty}$, then we find that the result does not match with the probability of the 
gravity--free case. Clearly, we are allowed to suppose that this second 
{\it Gedankenexperiment} takes place within the validity region of a LFCS of an adequately 
chosen curved manifold.

We may see that in the context of Feynman's formulation Ahluwalia's result seems to 
claim that the contribution to the corresponding probability coming from those 
trajectories that are not located within the validity region of LFCS is not the same as in the case in which we had no gravitational field. 
This conclusion clashes with that coming from the first {\it Gedankenexperiment}.

At this point there is an additional argument which could deserve a short remark. 
U\-sual\-ly, one of the doubts around the possibility of a consistent definition in the 
quantum realm of MCP concerns the fact that in QT physics is described by fields, 
which, of course, have a non--local 
character. This assertion is not very precise. Indeed, if 
we consider the description of a simple fluid in the context of GR, 
then we encounter the case of a system described also by fields 
(velocity field, pressure field, etc., etc.), 
which in consequence shares this non--local character, 
Nevertheless, the theory of Hydrodynamics in curved spacetimes [11] does not have the 
logical inconsistencies that beset QT in the context of GR.
    
If we take up the path integral formulation of QT, we may easily see [12] 
that the probability of finding a particle in a certain point depends on geometrical information 
that is associated to all trajectories that join these two points. Clearly, we can never 
find a finite neighborhood around any of these two points containing all these 
trajectories. It is readily seen that this last fact does not appear in the 
context of classical mechanics, in which we may always find a finite 
neighborhood, around the starting point or the final one, containing the whole needed information. 
In other words, in QT the geometrical information that renders the value of the probability in the final point 
has non--local character, and this nonlocality is incompatible with MCP, which is a postulate based on it. 
\bigskip
    
\section{Alternative definition of Minimal Coupling in Quantum Theory.}
\bigskip
    
A possible solution to this conceptual problem could be the modification of the integration domain in 
the corresponding path integral under the presence of a nonvanishing gravitational field. 
In other words, in order to evaluate the probability of finding our particle 
in $A$, 
knowing that it was previously in $P$, we could integrate only over those trajectories 
which lie completely 
inside the validity region of our locally flat coordinate system. This 
restriction 
would then allow us to evaluate the asked probability resorting only to those 
points of the corresponding 
manifold which lie completely within the validity region of LFCS. Of course, as was commented above, 
if we want to obtain, at least as a limit case, Feynman's propagator for a 
free particle, it seems also unavoidable that a weight functional has to be included.
     
Clearly, this weight functional must depend on the geometrical structures $G$ of the corresponding spacetime. 

Therefore we propose the following
\bigskip

\centerline{\bf Quantum Minimal Coupling Principle.}
\bigskip

The probability of finding a spinless particle in $A$, knowing that it was 
previously in $P$, is given by $|U_G|^2$
\bigskip

\begin{equation}
U_G(A, P) = \int_{\Omega}\omega_G[x(\tau)]d[x(\tau)]exp(iS[x(\tau)]/\hbar).
\end{equation}
\bigskip

Here $\Omega$ denotes the set of all trajectories joining $A$ and $P$, $S$ is the classical free particle action, 
$\omega$ is a weight functional that depends on the geometrical structures $G$ of the corresponding manifold and also on the trajectory (for instance, it is zero 
for those trajectories lying not completely within the validity region of the LFCS around $P$), and finally $\tau$ is any parametrization 
of the respective trajectories. Of course that in the absence of gravity $\omega_G[x(\tau)] =1$, for all trajectory.

As was mentioned before, this definition has the characteristic that in order to calculate the 
needed probability it suffices to have information about the geometrical structure of the corresponding manifold 
(it enters in the definition of $\omega_G[x(\tau)]$) and also the laws of Physics in Special Relativity. 
This non--local character of the required information in the context of QT implies that now we must know the 
geometry of the manifold in order to calculate this probability.

At this point it is noteworthy to mention how this new QMCP differs and at the same time coincides with the usual MCP. 
To begin, let us comment that if the probability coincides (under the adequate conditions) with Feynman's 
case, then we are in the spirit of MCP. Nevertheless, there is an additional 
point in which this proposal differs radically from the usual spirit of MCP, 
namely this principle claims that it suffices to have the laws of Special Relativity 
in order to know the result of any local experiment, i.e., it is unnecessary to 
have any kind of information concerning the geometry of the corresponding manifold. Here we have a different 
situation, because in this proposal we do need information about the involved geometry 
in order to know the result of any ``local'' quantum experiment. 

Let us now underline the mathematical similarity between this QMCP 
definition and RPIF [17], which is one of the possible formulations [18] that already exist in the context of DM, which tries to solve 
the so called quantum measurement problem. In other words, decoherence can be mathematically described in terms of 
RPIF [19].

This last remark allows us to interpret expression (1) stating that the geo\-me\-trical 
structures of our manifold act on particles as an always present measuring device, 
and in consequence it could render a geometrical explanation to the collapse of the wave function. 
This conclusion not only matches with an old claim: 
{\it gravity should play a fundamental role in approaches which could modify the formalism 
of QT} [20], but also connects the problem of the logical consistency of MCP in the context of QT with the old conundrum around the collapse of the wave function. 

On one hand this model coincides with several proposals 
that introduce the gra\-vi\-ta\-tional field as one of the physical entities that could give an explanation to this collapse [6, 7, 21, 22]. 
On the other hand, we must at this point also stress that this model has a fundamental difference with 
respect to these ideas which also use the gravitational field as an agent behind the collapse. 
If we take at look, for instance, at Diosi's work [7] we will immediately 
notice that in it the density operator acquires a stochastic behavior because 
the gravitational field does have fluctuations (with quantum origin) around the classical newtonian 
potential. In our case there are no spacetime fluctuations at all, and the gravitational field has a completely classical behavior.
\bigskip

\section{Free Particle Propagator.}
\bigskip

In this section we will calculate the propagator of a free particle in the context of the here proposed QMCP 
and recover (under the adequate conditions) Feynman's propagator [12].

The first problem that in this model we face concerns the correctness of the theo\-re\-tical 
predictions of RPIF. At this respect we must say that even though there are 
already theoretical results [23] which could render a feasible framework against the one 
these predictions could be confronted, the problem still remains open. The reason 
for this lies in the fact that those experiments that are required (for example the 
continuous measurement of the position of a particle in a Paul trap [24]) have not yet been carried out. 

The second problem is related to the choice of the involved weight functional 
appearing in expression (1). From RPIF we can not deduce the precise form of the correct weight 
functional, the exact expression depends on the measuring device [17]. In other words, in this 
particular case it depends on the gravitational field that we could have. 
But we do not know how a specific gravitational field could define its corresponding weight functional. 
Nevertheless, in a first approach we may accept a functional that could give the correct order of magnitude of the involved 
effects. Hence, knowing that in the first two cases in which this formalism 
was used the results coming from a Heaveside weight functional [25] and those 
coming from a gaussian one [26] coincide up to the order of magnitude, allows us 
to consider as our weight functional a gaussian one. This form has already been used 
to analyze the response of a gravitational wave antenna of Weber type [17], 
the measuring process of a gravitational wave in a laser--interferometer [27], 
or even to explain the emergence of the classical concept of time [28]. But a sounder justification 
of this choice comes from the fact that there are measuring processes in which the 
weight functional has precisely a gaussian form [29]. In consequence we could think about a curved manifold 
whose weight functional is very close to the gaussain form.

In order to simplify the calculation we will consider the case of a one--dimensional 
harmonic oscillator subject to the action of a gaussian weight functional. The res\-tric\-tion 
on the dimensionality of the system does not mean any lose of generality in our calculation, 
the reason for this stems from the fact that the general case can be obtained 
from the one--dimensional situation, we just have to multiply the one--dimensional case 
by itself three times. The condition of being a harmonic oscillator will disappear 
because we will consider the case of a harmonic oscillator which has vanishing frequency.

Therefore our starting point is the propagator of a particle with mass $m$ 
and frequency $w$
\bigskip

\begin{equation}
U_G(A, P) = \int_{\Omega}\omega_G[x(\tau)]d[x(\tau)]exp({i\over\hbar}S[x(\tau)]),
\end{equation}
\bigskip

where we have that
\bigskip

\begin{equation}
\omega_G[x(\tau)] = exp\{-{2\over T\Delta a^2}\int _{\tau '}^{\tau ''}[x(\tau) - a(\tau)]^2d\tau\}.
\end{equation}
\bigskip

Here $T = \tau '' - \tau'$ and $\Delta a$ represents the size of the validity region of LFCS.

At this point it is noteworthy to mention that the Weak Equivalence Principle (WEP) is still valid. Therefore in order to be able to recover it from expression (2) 
(which means that in the classical limit the motion of a free particle is given by geodesics) 
we have introduced in the weight functional the classical trajectory of the free case $a(\tau)$. 
Indeed, the classical behavior appears when $S/\hbar >>1$, and as we also 
know $S$ does not change, at least in first order, in the vicinity of the classical trajectory. 
Hence if $\omega_G[x(\tau)]$ changes much slower than the phase in the case 
$S/\hbar >>1$, then the main contribution to the propagator comes from an infinitesimal strip 
around the classical path (which in the case of a free particle is a geodesic). 
In consequence only the cla\-ssi\-cal trajectory has a nonvanishing probability, and in this way we recover WEP. 

The validity of WEP imposes a very strong 
condition on the set of possible weight functionals.
Indeed, only those $\omega_G[x(\tau)]$ whose rate of change is much slower 
than the rate of change of the phase (in the limit $S/\hbar >>1$) could be allowed. 
But this restriction yields also a theoretical argument against the one this QMCP could be 
confronted. If we could calculate the weight functional of any curved manifold, 
then in order to have the survival of WEP the resulting $\omega_G[x(\tau)]$ 
must fulfill this requirement, otherwise we would have manifolds in which WEP loses its validity.

 In expression (2) $S$ is the action of a harmonic oscillator
\bigskip

\begin{equation}
S[x(\tau)] = \int_{\tau '}^{\tau ''}L(x, \dot{x}),
\end{equation}
\bigskip

\begin{equation}
L(x, \dot{x}) = {1\over 2}m\dot{x}^2 - {1\over 2}mw^2x^2.
\end{equation}
\bigskip

This case is readily calculated [17]
\bigskip

\begin{equation}
U_G(A, P) = \sqrt{{m\tilde{w}\over 2\pi i\hbar sin(\tilde{w}T)}}
exp\Bigl(-2{<a^2>\over \Delta a^2} + {i\over\hbar}\tilde{S}_c\Bigr).
\end{equation}
\bigskip

In this last expression a few symbols need a short explanation. Here $\tilde{w}^2 = w^2 - i{4\hbar\over mT\Delta a^2}$, 
$<a^2> = {1\over T}\int_{\tau '}^{\tau ''}a^2(\tau)d\tau$, and finally $\tilde{S}_c$ is the classical 
action of the fictitious complex oscillator defined by $m\ddot{x} + m\tilde{w}^2x =0$.

Let us now consider the situation of a free particle, in other words, let us now take the case 
$w = 0$. Under this condition expression (6) becomes now
\bigskip

\begin{equation}
U_G(A, P) = \sqrt{{m\over 2\pi i\hbar T}{\sqrt{-i{4T\hbar\over m\Delta a^2}}\over sin(\sqrt{-i{4T\hbar\over m\Delta a^2}})}}
exp\Bigl(-2{<a^2>\over \Delta a^2} + {i\over\hbar}\tilde{S}_c\Bigr).
\end{equation}
\bigskip

Expression (7) is then the propagator of a free particle in this model. 
It is si\-mi\-lar 
to the propagator of a one--dimensional free particle whose coordinate is being continuously 
measured [17]. It contains an exponential damping term which depends on the ratio between $<a^2>$ 
and the size of the validity region of LFCS. If 
$<a^2>$ is much smaller than $\Delta a^2$ then damping plays no role at all in the dynamics 
of the particle.

Let us now recover Feynman's propagator and consider the limit 
$\sqrt{-i{4T\hbar\over m\Delta a^2}} \rightarrow 0$, which implies then that expression (7) becomes 
\bigskip

\begin{equation}
U_G(A, P) = \sqrt{{m\over 2\pi i\hbar T}}exp\Bigl({i\over\hbar}{m\over 2T}l^2)
exp\Bigl(-2{<a^2>\over \Delta a^2}\Bigr).
\end{equation}
\bigskip

Here $l$ is the distance between points $A$ and $P$. This imposed condition 
will be fulfilled if ${T\hbar\over m} << \Delta a^2$. 

Let us now analyze expression (8). The first two terms on the right hand side are 
identical to Feynman's free particle propagator. The last factor is a new contribution, has a damping character 
and is a direct consequence of the measuring role that in this QMCP play the degrees of freedom of the involved manifold. 
If we have also that $<a^2>/\Delta a^2 << 1$, then (8) becomes 

\begin{equation}
U_G(A, P) = \sqrt{{m\over 2\pi i\hbar T}}exp\Bigl({i\over\hbar}{m\over 2T}l^2).
\end{equation}
\bigskip

In order to estimate, at least very roughly, how good are (at points near to the Earth's surface) these two last approximations, let us 
take a weak field description for Earth's gravitational potential $\phi$ near its surface. 
Under this condition we have that [11] $R^j_{oko} \sim {\partial ^2\phi\over\partial x^j\partial x^k}$, which implies $\Delta a \sim 10^{13}$ cm. 
(this is no surprise at all, indeed if we consider a different but related case, 
namely the size of the validity region of the coordinate system of a uniformly accelerated observer whose acceleration is equal to the 
magnitude of gravity on Earth's surface, then we find that this region has a 
size of approximately 1 light--year $\sim 10^{18}$ cm. [11]). 
Then for a free electron ${T\hbar\over m} << \Delta a^2$ breaks down if $T \sim 10^{25}$ sec., and $<a^2>/\Delta a^2 << 1$ is not anymore fulfilled when $l \sim 10^{12}$ cm. In other words, on Earth's surface the here proposed model can not be distinguished 
from Feynman's case.

\bigskip

\section{Diatomic Interstellar Molecules.}
\bigskip

An interesting case could be the analysis in this model of the movement of simple interstellar molecules. 
The reason for this stems from the fact that recently it was claimed that QT could 
participate in the determination of the structure and size of galaxies [30]. 
Therefore we may wonder how the movement of a simple in\-ter\-ste\-llar molecule looks like in this proposal. 
The idea here is to comprehend better the differences (with respect to the usual case) of 
the behaviour of this kind of matter and see if some new effects could emerge.

It is already known that the so called {\it giant molecular clouds} are an important component 
of the interstellar medium [31]. These clouds are the coolest 
components of it, temperatures are in the range 10 up to 100 K, and 
contain several diatomic molecules, for instance, CO, CH, CN, CS, or C$_2$ [31].

In the case of a diatomic molecule the ``effective'' Hamiltonian contains 
a potential term $V(R)$ (where $R$ is the separation between the nuclei) which 
includes not only the Coulomb repulsion of the nuclei but also the effective potential due 
to the electron configuration. This attractive potential can be approximated, for small values of $R$, 
with a linear oscillator. This approximation is much better for heavy molecules (better for a Xe--Xe molecule than for a Ne--Ne molecule) [32].    
A better description is obtained by means of a Lennard--Jones potential.
 
If we consider only the vibrational description for the nuclei (rotational degrees of freedom are neglected), then we may reduce the whole problem to the analysis of a harmonic oscillator, at least in a first approach. 

If one of these diatomic molecules is located in a region in which a nonvanishing 
gravitational field exists, then the propagator of its associated harmonic oscillator can be approximated with expression (6).

The relative probability for these systems is in the case of large $T$ (${4\hbar\over mTw^2\Delta a^2} << 1$) 

\begin{equation}
P = {mw\over 2\pi\hbar sin(wT)}exp\Bigl(-4{<a^2>\over\Delta a^2}~\Bigr)
\end{equation}
\bigskip

Clearly, the wave function shows a spreading with time which does not appear in the usual case, 
i.e., ${mw\over 2\pi\hbar sin(wT)}$. Therefore it seems that in a first approach some simple diatomic 
interstellar molecules would not be so strongly localized as in the common situation. 
We may wonder if this new spreading of the wave function of some components of the interstellar matter could imply some change in the current models that seek an explanation for 
the appearance of cosmic structures. 

Assume now the limit ${4\hbar\over mTw^2\Delta a^2} >> 1$. Hence the relative probability for this system becomes
\bigskip

\begin{equation}
P = {mw\over 2\pi\hbar}\sqrt{{2\over cosh\Bigl(2\sqrt{{2\hbar T\over m\Delta a^2}}~\Bigr) - cos\Bigl(2\sqrt{{2\hbar T\over m\Delta a^2}}~\Bigr)}}
exp\Bigl(-4{<a^2>\over\Delta a^2}~\Bigr)
\end{equation}
\bigskip

Here the relative probability diminishes, not only as a consequence 
of the purely geometrical term $exp\Bigl(-4{<a^2>\over\Delta a^2}\Bigr)$, but also 
as a consequence of time $T$. 
\bigskip

\section{Conclusions.}
\bigskip

Ahluwalia has pointed out that the general--relativistic description of gravitation at the quantum realm 
can not be considered complete.  
In this work we have tried to show that the geometrical information that we need 
to calculate probabilities (\'a la Feynman) in any point 
of a curved manifold is not enclosed in a region with finite volume, a non--local characteristic. 
Gravity--induced non--locality has already been analyzed [33] and can be interpreted 
[34] in the context of spinors as a gravity--induced CP violation which renders 
a dynamical explanation to the collapse of a neutron star into a black hole and to the involved loose of information. 
Therefore a more profound analysis of this gravity--induced non--locality could be important.

In relation with this non--locality of QT we have introduced a new QMCP, which comprises two important differences with respect to the usual 
standpoint in GR, namely;

(1) We must now abandon an old postulate, which states: the probability of finding a particle in $A$, knowing 
that it was previously in $P$ (both points within the validity region of LFCS), 
can be calculated without having any kind of information about the geometrical structure 
of the corresponding manifold.

(2) The second difference concerns the introduction of a restriction in the integration domain of the corresponding path integral. 
Mathematically this restriction is expressed by means of a weight functional, which would contain information about the geometrical 
structure of the involved particle.
    
The price paid, in connection with the need of knowing the information about 
the geometry of spacetime, allows us to build a bridge which establishes a relation between 
two topics that up to now have been not related, namely DM and QMCP. In this proposal the degrees of freedom of geometry play the role of a measuring device 
which acts always on a quantum particle. This idea coincides, at least partially, with a 
claim stating that the gravitational field could be one of the physical entities behind the collapse of the wave 
function.

Finally, we calculated the propagator of a free particle and have also found 
that it contains, under the appropriate conditions, Feynman's result. The difference 
comprises an exponentially decaying factor, which depends only on the 
length of the cla\-ssi\-cal trajectory and on the size of the validity region of LFCS. Only if the distance of the displacement has an order of magnitude 
similar to $\Delta a^2$ would the damping term appear in scene. 
This new contribution to the pro\-pa\-ga\-tor depends only on geometrical parameters (no dependence on the mass), 
and in consequence is the same for all kind of particles. We have also shown that near the Earth's surface 
the new effects that in this model appear are completely negligible. In other words, this proposal coincides 
in any terrestrial experiment with the usual predictions.

As was commented in section three, in this model gravity could drive a collapse 
of the wave function (because geometry acts as a measuring device) but the role that it plays is not the same as in 
other models that in this direction already exist. Indeed, if we take a look at Diosi's work [7] we may see that 
the density operator acquires a stochastic behavior, which stems from 
fluctuations (with quantum origin) of the newtonian gravitational field. 

An approximated expression for the propagator of the harmonic oscillator a\-sso\-cia\-ted 
to a simple diatomic molecule situated in interstellar space has also been derived. 
It has been proved that this case shows a spreading with time, which emerges as a consequence of the measuring role that 
geometry plays in this model.

Let us also mention that the description of the problem of measurement in 
QT 
has at least five different approaches, which are mathematically equivalent [18]. 
One of them is precisely RPIF, and at this point we may wonder if we might 
express this new QMCP in the formalism of the group approach to the master equation [35], 
as a nonlinear stochastic differential equation [36], or even in terms of the remaining 
two mathematical models.

Some points that in the here proposed context could be interesting to analyze 
are the introduction of spin in expression (1), this could allow us to deduce the incompleteness inequality 
that in connection with flavor--oscillation clocks appears in [5], as well as the generalization 
of (1) in order to include the case of Dirac's equation.  

It is noteworthy to mention that we have introduced a modification in QT, which has its origin in the degrees of freedom of 
geometry, and therefore could shed some light on the problem of a quantum theory 
of gravity. Indeed, there are some claims [37] which state that not only GR has to be modified in order to have 
a quantum theory of GR, but also that QT has to suffer modifications.
\bigskip

\Large{\bf Acknowledgments.}\normalsize
\bigskip

The author would like to thank A. Camacho--Galv\'an and A. A. Cuevas--Sosa for their 
help, and D.-E. Liebscher for the fruitful discussions on the subject. 
The hospitality of the Astrophysikalisches Institut Potsdam is also kindly acknowledged. 
This work was supported by CONACYT Posdoctoral Grant No. 983023.

\end{document}